%% file: main_final_version.tex
\documentclass[lettersize,journal]{IEEEtran}
\usepackage{amsmath,amsfonts}
\usepackage{booktabs}
\usepackage{arydshln}
\usepackage{algorithmic}
\usepackage{algorithm}
\usepackage{array}
\usepackage[caption=false,font=normalsize,labelfont=sf,textfont=sf]{subfig}
\usepackage{textcomp}
\usepackage{stfloats}
\usepackage{url}
\usepackage{verbatim}
\usepackage{graphicx}
\usepackage{multirow}
\usepackage{booktabs}
\usepackage{xcolor}
\usepackage{cite}
\usepackage{comment}
\newcommand{\be}{\begin{eqnarray}}
\newcommand{\ee}{\end{eqnarray}}
\begin{document}

\title{Speech Dereverberation with Frequency Domain Autoregressive Modeling}

\author{Anurenjan Purushothaman,~\IEEEmembership{Member,~IEEE}, Debottam Dutta, Rohit Kumar,~\IEEEmembership{Student Member,~IEEE} and Sriram Ganapathy,~\IEEEmembership{Senior Member,~IEEE}
        % <-this % stops a space
\thanks{This paper was partly funded by grants from the Samsung Research India, Bangalore.}% <-this % stops a space
\thanks{A. Purushothaman, and S. Ganapathy are with the Learning and Extraction of Acoustic Patterns (LEAP) lab, Department of Electrical Engineering, Indian Institute of Science, Bangalore, India, 560012. D. dutta is with University of Illinois Urbana-Champaign. R. Kumar is   associated with Johns Hopkins University. 
 e-mail: \{anurenjanp, sriramg\}@iisc.ac.in}} 

\markboth{Journal of \LaTeX\ Class Files,~Vol.~xx, No.~xx, January~2023}%
{Shell \MakeLowercase{\textit{et al.}}: Speech enhancement with frequency domain auto-regressive modeling}


\maketitle

\begin{abstract}
Speech applications in far-field real world settings often deal with  signals that are corrupted by reverberation. The task of dereverberation  constitutes an important step to improve the audible quality and to reduce the error rates in applications like automatic speech recognition (ASR). 
We propose a unified framework of speech dereverberation for improving the speech quality and the ASR performance using the approach of envelope-carrier decomposition provided by an autoregressive (AR) model. The AR model is applied in the frequency domain of the sub-band speech signals to separate the envelope and carrier parts. A novel neural architecture based on dual path long short term memory (DPLSTM) model is proposed, which jointly enhances the sub-band envelope and carrier components.  The  dereverberated envelope-carrier signals are modulated and the sub-band signals are synthesized to reconstruct the audio signal back. The DPLSTM model for dereverberation of envelope and carrier components also allows the joint learning of the network weights for the down stream ASR task.    
In the ASR tasks on the REVERB challenge dataset as well as on the VOiCES dataset, we illustrate that the joint learning of speech dereverberation network and the E2E ASR model yields significant performance improvements over the baseline ASR system trained on log-mel spectrogram as well as other benchmarks for dereverberation (average relative improvements of $10$-$24$\% over the baseline system). The speech quality improvements, evaluated using subjective listening tests, further highlight the improved quality of the reconstructed audio. 
\end{abstract}


\section{Introduction}
 
\IEEEPARstart{T}{he} wide spread adoption of voice technologies like  meeting assistants, smart speakers, in-car entertainment systems, and virtual assistants imply that the audio signal at the input of these system is impacted by reverberation and noise artifacts \cite{haeb2020far}.  The performance of the downstream applications like, automatic speech recognition, speaker/language recognition, emotion recognition or voice activity detection, is shown to degrade significantly in reverberant conditions \cite{hain2012transcribing, dan, ganapathy20183, gusev2020deep, 4291604}. The performance deterioration is primarily attributed to the smearing of the temporal envelopes caused by reverberation \cite{yoshioka2012making}.  The temporal smearing is caused by the emplacement of the direct path signal on reflected signals, resulting in a weighted summation of delayed components~\cite{ganapathy2012signal}.

One of the approaches to deal with the adverse far-field conditions is to develop a front-end which performs signal enhancement. Several techniques for dereverberation like signal processing based (for example, weighted prediction error (WPE) \cite{wpe}), mask estimation based (for example, time-frequency mask estimation \cite{williamson2017time}) and multi-channel beamforming based (for example, time-delay estimation \cite{anguera2007acoustic}, generalized eigen-value \cite{warsitz2007blind, rohit}) have been explored to improve the signal quality. On the other hand, another effective approach for system development in reverberant conditions is that of multi-condition training  \cite{seltzer2013investigation}. 
%Here, either simulated or real far-field data is used to the train the models. 
However, even with these pre-processing and multi-condition training methods, the beamformed signal contains significant amount of temporal smoothing which adversely impacts the ASR performance~\cite{peddinti2017low}.

\textcolor{black}{In the traditional setting, the first step in the analysis of a signal is the short-term Fourier transform (STFT). The key assumptions about the convolution model of reverberation artifacts, is applicable  for a long-analysis window in the time domain,
\textcolor{black}{or using convolutional transfer function with cross-band filters in the STFT domain \cite{talmon2009relative,avargel2007system}.  In our case, we use the former approach of long analysis window and explore dereverberation in the sub-band envelope domain}. As the reverberation is a long-term convolution effect, we highlight that room impulse response (typically with a T$60 > 400ms$)  can be absorbed as a multiplication in the frequency domain, as well as a convolution in the sub-band envelope domain.} 

In this paper, we investigate the effect of reverberation on the long-term sub-band signals of speech using an envelope-carrier decomposition. 
The extraction of the sub-band envelope is achieved using the autoregressive (AR) modeling approach in the spectral domain, termed as frequency domain linear prediction (FDLP).
Our previous work showed that a feature level enhancement with the FDLP envelope improves speech recognition performance  \cite{purushothaman2022dereverberation, purushothaman20203}. However, the prior works did not allow the reconstruction of the audio signal for quality improvement. Further, the enhancement of the carrier signal was not addressed in the previous work due to the challenges in the handling the impulsive nature of the carrier signal. 

In this paper, we propose a novel approach to the joint dereverberation of the envelope and carrier signals using a neural modeling framework.
\textcolor{black}{While using the sub-band signals directly, the sample level de-convolution with a suitable loss function can be a difficult design choice to learn using neural models. Hence, we propose using an envelope-carrier decomposition of the sub-band signals. Our rationale for the envelope-carrier decomposition based setup is the fact the envelope information is alone used in the ASR experiments. Thus, the ASR loss has to  impact only the envelope dereverberation branch.  However, the carrier and the envelope components are part of the signal reconstruction branch. 
%We hypothesize that this segregation of the streams into envelope and carrier is important for the dereverberation task catered to ASR and signal reconstruction applications.
}

We develop a dual path long short term memory (DPLSTM) architecture for the dereverberation of the temporal envelope and carrier signals. \textcolor{black}{In our case, the goal of the neural model is to perform a dereverberation of the envelope and the carrier components of the sub-band signal. These signals have a time profile, with varying dynamic range and properties.  Further, merging all the sub-band signals in the decomposition also brings in a frequency profile. Thus, the design choice of the neural model, for enhancing the sub-band envelope-carrier signals, has to learn the sequence level patterns in both the time and frequency domains. The DPLSTM \cite{dprnn} is a suitable choice, as the model is able to integrate information effectively in both the time and frequency domains.}

Following the dereverberation step, the sub-band modulation and synthesis step generates the reconstructed audio signal. The neural enhancement and sub-band synthesis can also be implemented as a part of the larger neural pipeline for downstream tasks like ASR, thereby enabling the joint learning of the ASR and dereverberation model parameters. We refer to the proposed approach as Dual path dereverberation using Frequency domain Auto-Regressive modeling (DFAR) and the joint end-to-end model as E2E-DFAR. 

Various ASR experiments are performed on the REVERB challenge dataset \cite{reverb} as well as the VOiCES dataset \cite{voices,voices_inter}. 
%In these experiments, we show that the E2E-DFAR method improves over the state-of-the-art speech enhancement systems and ASR systems for the respective tasks.
The key contributions from this work, over the prior work \cite{purushothaman2022dereverberation}, can be summarized as follows, 
\begin{itemize}
    \item Proposing an analysis for dereverberation with a sub-band decomposition  and envelope-carrier demodulation. 
    \item Proposing a dual-path long short time memory model named, DPLSTM for the dereverberation of sub-band envelope and carrier signals. This approach is termed as DFAR. 
    \item Developing a joint learning scheme, where the ASR model and the DFAR model are optimized in a single end-to-end framework. This model is referred to as the E2E-DFAR.
    \item Evaluating the proposed approaches on speech quality improvement tasks as well as on ASR tasks on two benchmark datasets - REVERB challenge dataset and the VOiCES dataset. 
\end{itemize}

%The rest of the paper is organized as follows. The related prior work is discussed in Section~\ref{sec:prior_work}.  Section~\ref{sec:proposed model} provides details regarding the proposed approach. The experimental set up is detailed in section~\ref{sec:exp_set_up}. The experiments and results are discussed in Section~\ref{sec:expt}. This is followed by a conclusion of the work in Section~\ref{sec:summary}.

\section{Related prior work}\label{sec:prior_work}
%In this section, we highlight the recent works in the area of speech enhancement and dereverberation as well as methods proposed for robust ASR in far-field conditions. 

\subsection{Enhancement and dereverberation}
For speech enhancement, Xu et. al. \cite{xu2014regression} devised a mapping from noisy speech to clean speech using a supervised neural network. In a similar manner, ideal ratio mask based neural mappings  \cite{wang2018supervised} have been explored for speech separation tasks. On the dereverberation front, Zhao et. al. proposed an LSTM model for late reflection prediction in the spectrogram domain  \cite{zhao2018late}. Han et. al \cite{han2014learning} developed a spectral mapping approach using the log-magnitude inputs and Williamson et. al \cite{williamson2017time} proposed a mask-based approach for dereverberation on the complex short-term Fourier transform. In a different line of work, speech enhancement in the time domain was pursued by Pandey et. al \cite{pandey_2019}.   

The application of speech dereverberation as a pre-processing step for downstream applications like ASR have been explored in several works (for example, \cite{wollmer2013feature,chen2015speech,weninger2015speech}). The recent years have seen the use of recurrent neural network architectures for dereverberation.  For example, Maas et. al \cite{maas2013recurrent}, utilized a recurrent neural network (RNN) to establish mapping  between noise-corrupted  input  features  and  their  corresponding  clean  targets. Also, the use of a context-aware recurrent neural network-based convolutional encoder-decoder architecture was investigated by Santos et. al. \cite{santos2018speech}. 
\begin{figure*}[t!]
  \centering
  \includegraphics[scale=0.18]{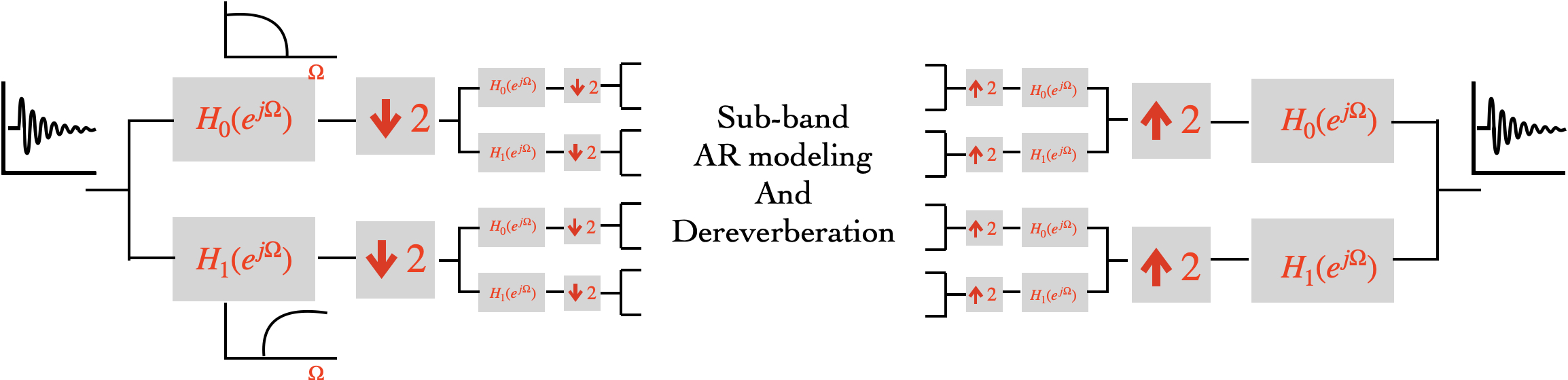}
\caption{Illustration of a 4-channel uniform QMF decomposition using a 2-stage binary QMF tree. In our work, we use $64$-channel decomposition, using a 6-way binary tree. }
  \label{fig:QMF-decomposition}
  \vspace{-0.1in}
\end{figure*}
\subsection{Robust multi-channel ASR}
In the design of robust ASR, Generalized sidelobe canceller (GSC) \cite{griffiths1982alternative,gannot2001signal} is a common approach. It was introduced by  Li et. al in \cite{li2021deep}, where the authors  proposed a neural network-based generalized side-lobe canceller. To combine spectral and spatial information from multiple channels using attention layers, an end-to-end multi-channel transformer was investigated in \cite{chang2021end}. In another attention modelling approach, the streaming ASR model based on monotonic chunk-wise attention was  proposed by Kim et. al in \cite{kim2021streaming}.  Ganapathy et. al. \cite{ganapathy20183} proposed a 3-D CNN model for far-field ASR.
%, where the data from  all the microphones are chosen as input  to the ASR system without a beamforming step. 

\subsection{Joint modeling of enhancement and ASR}
The attempt  proposed by Wang et. al. \cite{wang2016joint} incorporates a DNN based speech separation model coupled with a DNN based acoustic model. The work reported by Wu et. al. \cite{bowu2e2017} explored a unification of separately trained speech enhancement neural model and the acoustic model, where the joint model is fine-tuned to improve the ASR performance. Here, the DNN based dereverberation front end leverages the knowledge about reverberation time. 
While traditional GSC is optimized for signal level criteria, the neural network-based GSC, proposed by Li et. al \cite{li2021deep}, was optimized for ASR cost function.

\section{Proposed DFAR approach}\label{sec:proposed model}
%In this section, we provide the details of the sub-band decomposition, frequency domain linear prediction and dual-path LSTM model, all of which form parts of the proposed DFAR model.  

\subsection{Quadrature Mirror Filter (QMF)}
% \begin{figure*}[t!]
%   \centering
%   \includegraphics[scale=.4]{qmf.pdf}
% \caption{The 64 channel uniform QMF derived using 6-stage network. Input signal
% $x[n]$ is sampled at 16 kHz. $LP$ and $HP$ denote Low-Pass and High-Pass band, respectively. $\downarrow 2$ denotes down-sampling by 2. $x_q[n]$ denotes the $q^{th}$ sub-band signal.}
%   \label{fig1}
% \end{figure*}
\textcolor{black}{
For the sub-band decomposition, we had the following design considerations
\begin{itemize}
    \item  The decomposition approach should allow the long-term artifacts of reverberation to be captured in the sub-band domain as a convolution, 
    \item The analysis method should allow a perfect reconstruction back to the audio using the synthesis part, and 
    \item The sub-band components should be critically sampled for efficient computation of the dereverberated components in a deep neural model. 
\end{itemize}
The quadrature mirror filter (QMF) met all the above requirements and hence, this work has used the QMF analysis and synthesis for speech dereverberation task.}

\textcolor{black}{A quadrature mirror filter (QMF) is a filter whose magnitude response is a mirror reflection at quadrature frequency ($\frac{\pi}{2}$) of another filter \cite{vaidyanathan1987quadrature}}. In signal processing, the QMF filter-pairs are used for the design of perfect reconstruction filter banks. Let $H_0(e^{j\Omega})$ and $H_1(e^{j\Omega})$ denote low-pass and high-pass filter's  frequency domain function, where $\Omega$ is the digital frequency. In addition to the quadrature property ($H_1(e^{j\Omega}) = H_0(e^{j(\Omega - \pi)})$), the filters used in QMF filter-banks also satisfy the complimentary property, 
\be
|H_0(e^{j\Omega})|^2 + |H_1(e^{j\Omega})|^2 = 1.
\ee

The design of sub-band decomposition scheme with QMF involves a series of filtering and down-sampling operations for the analysis \cite{vaidyanathan2006multirate}. The synthesis is achieved by up-sampling and filtering operations. A tree-like structure can be formed using a recursive decomposition operation. The down-sampling process enables a critical rate of processing, where the sum of the number of samples in each sub-band equals the number of the samples in the full-band signal.  

In this work, we use an uniform $64$-band Quadrature Mirror Filter bank (QMF) for decomposing the input signal into $64$ uniformly spaced frequency bands. Inspired by the audio decomposition scheme outlined in Motlicek et. al.  \cite{motlicek2007scalable}, we use a 6-level binary tree structure. The schematic of the sub-band decomposition is shown in Fig.~\ref{fig:QMF-decomposition}. For the implementation in a neural pipeline, the down-sampling operation is equivalent to a stride, while the up-sampling operation is that of un-pooling.  

\subsection{Autoregressive modeling of temporal envelopes}
The application of linear prediction model in the frequency domain, an approach called frequency domain linear prediction (FDLP), enables the modeling of the temporal envelopes of a signal with an autoregressive (AR) model \cite{athineos2003frequency,ganapathy2012signal}. 
\textcolor{black}{The sub-band signal is transformed to the spectral domain using a discrete cosine transform (DCT) \cite{ganapathy2012signal}, where a linear prediction model is applied}. 

Let the sub-band signal be denoted as $x_q[n]$, where  $q=1,..,Q$ denotes the sub-band index.
\textcolor{black}{The analytic signal, in signal processing theory, is a complex valued function, whose real value is the original signal while the imaginary value is the Hilbert transform of the signal. It finds application in single side-band amplitude modulation and quadrature filtering.}
Let the analytic version of sub-band signal, $x_q[n]$ be denoted as, $x_q^a[n]$. The corresponding analytic signal in the frequency domain, $X_q^a[k]$ can be shown to be the one-sided discrete Fourier transform (DFT)~\cite{ganapathy2012signal} of the even symmetric version of $x_q[n]$.

We apply linear prediction (LP) on the frequency domain signal, $X_q^a[k]$. The corresponding LP coefficients are denoted by $\{b_p\}_{p=0}^m$, where $m$ is the order of the LP.  The temporal envelope estimate of $x_q^a[n]$, is given by, 
\begin{eqnarray}\label{eq:envelope_lp_model}
e_q[n] = \frac {\alpha}{|\sum_{p=0}^m b_p e^{-2 \pi i p n} |^2} 
\end{eqnarray} 
where $\alpha$ denotes the LP gain. The envelope represents the autoregressive model of the Hilbert envelope. In this paper, we use the Burg method \cite{burg1975maximum} for estimating the AR envelope. 

The corresponding carrier (remaining residual signal), $c_q[n]$ is found as, 
%by sample wise division of the signal $x_q[n]$ by the estimated envelope $e_q[n]$.
\begin{eqnarray}
c_q[n] = \frac{x_q[n]}{\sqrt {e_q[n]}}
\end{eqnarray}
The division operation in the expression above is well defined as the envelope given in Eq.~(\ref{eq:envelope_lp_model}) is always positive. Further, the modeling of the temporal envelopes using the AR model ensures that the peaks of the sub-band signal in the time-domain are well represented \cite{ganapathy2009autoregressive,ganapathy2014robust}. 

\subsection{Effect of reverberation on envelope and carrier signals}
The effect of reverberation on the time-domain speech signal can be expressed in the form of a convolution operation, 
\be
\label{eq:reverb_sig}
y[n] = x[n]*r[n],
\ee
where $x[n]$ denotes the clean speech signal, $r[n]$ is the impulse response of the room and $y[n]$, is the reverberant speech signal. The room response function can be further split into two parts, $r[n] = r_e[n] + r_l[n]$, where $r_e[n]$ and $r_l[n]$ are the early and late reflection components, respectively. 

Let $x_q[n]$, $r_q[n]$ and $y_q[n]$ denote the sub-band versions of the clean speech, room-response function and the reverberant speech signal respectively.  Assuming an ideal band-pass filtering,  it can be shown that the analytic signal, $x_q^a[n]$, is given by \cite{thomas,ganapathy2012signal},
\be
\label{eq:reverb_analytic_conv}
y_q^a[n] = \frac{1}{2}[x_q^a[n]*r_q^a[n]],
\ee
For band-pass filters with narrow band-width, the envelopes of the reverberant speech can  be approximated as \cite{purushothaman2022dereverberation},
\be
\label{eq:envelope_conv_model}
e_{yq}[n] \simeq \frac{1}{2} e_{xq}[n]*e_{rq}[n],
\ee
where $e_{yq}[n]$, $e_{xq}[n]$, $e_{rq}[n]$ denote the sub-band envelopes of reverberant speech, clean speech and room response respectively. \textcolor{black}{Prior efforts in envelope normalization focus on suppressing the linear effects of reverberation by setting the gain of the reconstructed envelopes to unity \cite{ganapathy2011feature}}. However, in this work, we develop neural models that can remove the non-linear effects of reverberation. The reverberant sub-band envelope can also be viewed an additive model \cite{purushothaman2022dereverberation, kumar2022end}.
\be
\label{eq:envelope_conv_model_early}
\textcolor{black}{e_{yq}[n] = e_{yqe}[n] + e_{yql}[n]},
\ee
where, \textcolor{black}{$e_{yqe}[n]$ is the early reflection component (which includes the direct path and the early reflections),} while $e_{yql}[n]$ is the late reflection part of the   sub-band envelope $e_{yq}[n]$.

\textcolor{black}{The key assumptions about the reverberation model of  Eq. (\ref{eq:reverb_sig}-\ref{eq:envelope_conv_model}), is a long-analysis window in the time domain. As the reverberation is a long-term convolution effect, we highlight that the room impulse response (typically with a T$60 > 400ms$)  can be absorbed as a multiplication in the frequency domain, as well as a convolution in the sub-band envelope domain, only in the case of a long analysis window. The widely used short-time Fourier transform (STFT) does not capture the room impulse response function directly, and hence does not allow a convolutive modeling of the artifacts. Further, the phase effects in STFT domain are somewhat cumbersome to model. The above mentioned issues of STFT are also verified experimentally in Sec.~\ref{sec:expt}.}
\begin{figure*}[t!]
  \centering
  \includegraphics[scale=.32]{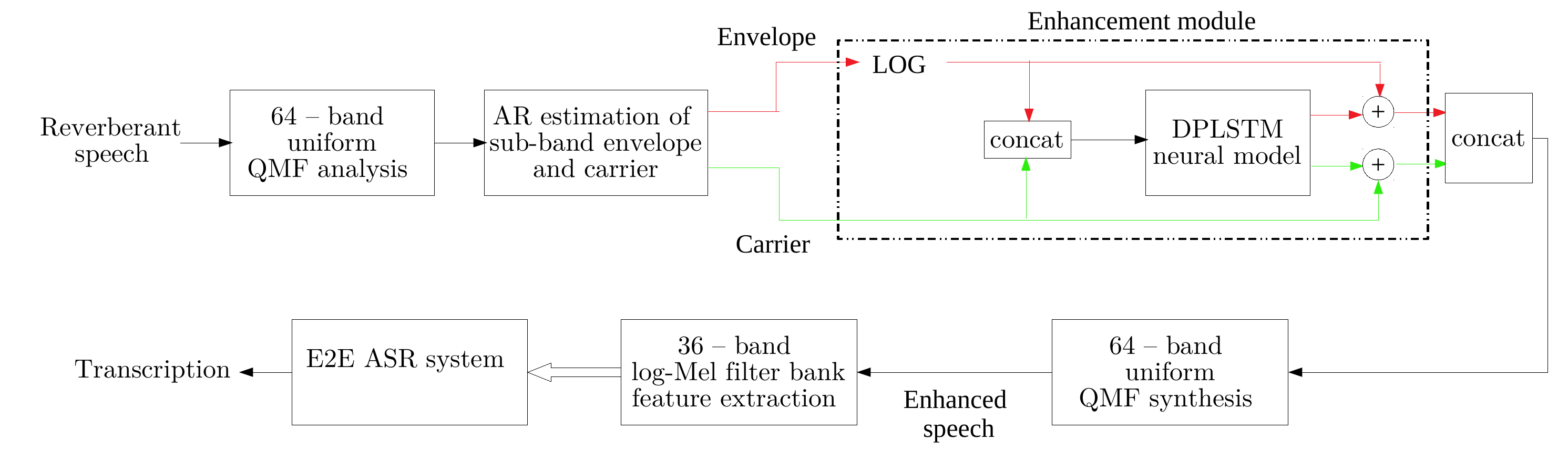}
\caption{Block schematic of speech dereverberation model, the feature extraction module and the E2E ASR model. The red arrows denote the envelopes, $e[n]$, and the green arrows represent the carrier, $c[n]$. The entire model can be constructed as an end-to-end neural framework.}
  \label{fig:joint-model}
  \vspace{-0.1in}

\end{figure*}

\textbf{Envelope enhancement:} A neural model can be used to learn late reflection component $e_{xql}[n]$ from the sub-band temporal envelope $e_{xq}[n]$. The predicted late reflection component can be 
subtracted from the sub-band envelope to suppress the artifacts of reverberation. 
%This approach is spectral subtraction approach in speech enhancement \cite{martin2005speech}. 

We pose the problem in the log domain to reduce the dynamic range of the envelope magnitude. The neural model is trained with reverberant sub-band envelopes ($log~(e_{xq}[n])$) as input.  The model outputs the gain (in the log domain, i.e., $log~\frac {e_{sq}[n]}{e_{xq}[n]}$). This gain is added in the log-domain to generate dereverberated signal envelope ($log~(\hat{e}_{sq}[n]$). 
\begin{figure*}[t!]
  \centering
  \includegraphics[scale=0.14]{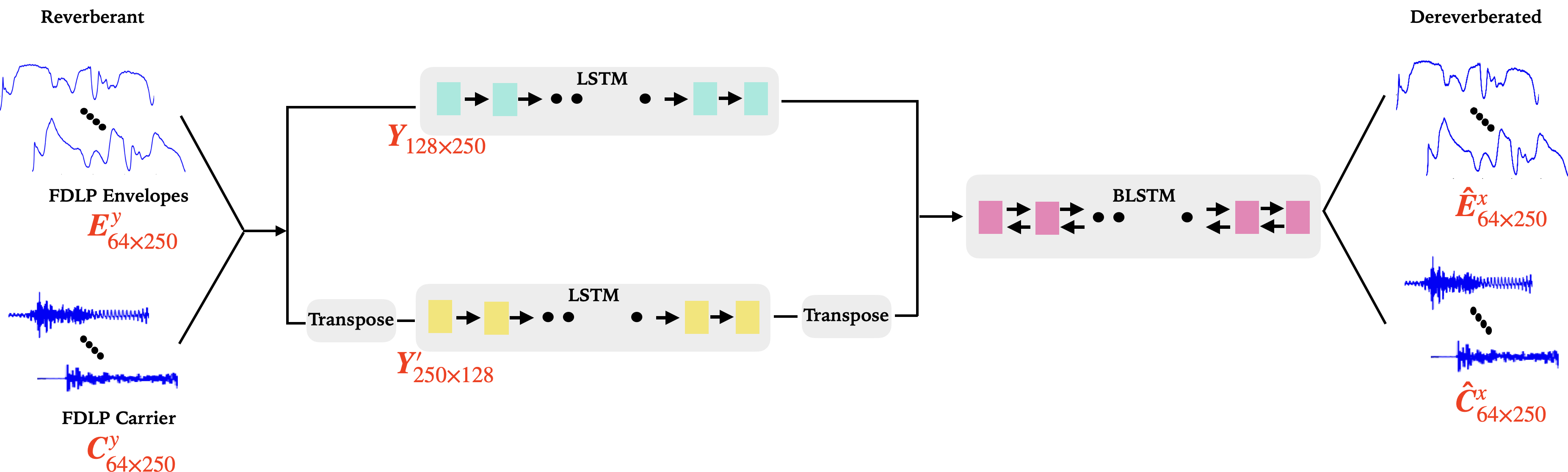}
    \caption{The dual path LSTM model architecture for envelope-carrier dereverberation. The top LSTM path   models the recurrence along the time dimension while the one on the bottom models the recurrence along the frequency dimension.}
  \label{fig_dplstm}
  \vspace{-0.1in}

\end{figure*}

 \textbf{Envelope-carrier dereverberation model}: In a similar manner, 
 %we hypothesize that reverberant carrier of the sub-band signal, $c_{xq}[n]$ follows an additive model as,
%\textbf{Carrier enhancement}:
the non-linear mapping between the reverberant carrier, $c_{xq}[n]$ and clean carrier, $c_{xq}[n]$, can be learned using a neural network. A neural model is trained with reverberant sub-band carrier ($c_{xq}[n]$) as input and model outputs the residual (an estimate of the late reflection component, $c_{xql}[n]$), which when added  with the reverberant carrier generates the estimate of source signal carrier ($\hat{c}_{sq}[n]$).  Instead of independent operations of dereverberation of the envelope and the carrier, we propose to learn the mapping between clean and reverberant versions of both the envelope and the carrier in a joint model. The input to the neural model is the sub-band reverberant envelope spliced with the corresponding carrier signal. The network is trained to output the late reflection components of both the envelope and carrier. With this approach, the model also learns the non-linear relationships between the envelope and carrier signals for the dereverberation task. 
%The late reflection component of envelope is in the log domain, $log~(e_{xq}[n]$ and late reflection component of carrier is $c_{xql}[n]$. 
From the model output,  the estimate of the clean sub-band signal $\hat{s}_q[n]$ is generated. In our implementation, the audio signal is divided into non overlapping segments of $1$ sec. length and  passed through the envelope-carrier dereverberation model. The model is outlined in Fig.~\ref{fig:joint-model}.

\subsection{DFAR model architecture using DPLSTM}
We propose the dual path long short term model (DPLSTM) for the dereverberation of the envelope-carrier components of the sub-band signal. 
 Our proposed model is inspired by dual path RNN  proposed by Luo {et. al} \cite{dprnn}. The block schematic of the DPLSTM model architecture is shown in Fig.~\ref{fig_dplstm}. For $1$ sec. of audio sampled at $16$ kHz, the envelope ($\mathrm{\boldsymbol{E}}^y$) and carrier ($\mathrm{\boldsymbol{C}}^y$) components of the critically sampled sub-band signals ($64$ channel QMF decomposition) are of length $250$. The  envelope/carrier signals of all the sub-bands, for the reverberant signal ($\mathrm{\boldsymbol{Y}}$), is  of size $64 \times 250$. The combined envelope-carrier input is therefore of size $128 \times 250$, which forms the input to the DPLSTM model. The DPLSTM model outputs are also of the same size of the input, and the model is trained using the mean squared error (MSE) loss.

The proposed DPLSTM has two paths, one LSTM path models the recurrence along the time dimension, while the other models the recurrence along the frequency dimension. We use two separate $3$-layer LSTM architectures for these paths. The output dimensions are kept the same as the input dimension for each of these paths. The frequency recurrence LSTM output is transposed and these are concatenated in the frequency dimension. This combined output is fed to a multi layer bi-directional LSTM, which performs recurrence over time. The final output is split into sub-band specific envelope and carrier components. The modulation of the envelope with the respective carrier components generates the sub-band signals, which are passed through the QMF synthesis to generate the full-band dereverberated signal.  

\subsection{Joint learning of dereverberation model  for ASR}
The joint learning of the  envelope-carrier dereverberation module with the E2E ASR architecture is achieved by combining the two separate models to train a single joint neural model. This is shown in Fig.~\ref{fig:joint-model}.  We initialize the modules with weights obtained from the independent training of each component. Specifically, the envelope-carrier dereverberation model is trained using MSE loss, which is followed by a sub-band synthesis (right side half of Fig.~\ref{fig:QMF-decomposition}). The QMF synthesis is implemented using a 1-D CNN layer to generate the dereverberated speech signal. Further, the E2E ASR architecture is  separately trained on the log-mel filter bank features, obtained from the dereverberated speech. The mel-filter bank feature generation can also implemented using a neural framework. Thus, the final model, composed of neural components from the envelope-carrier dereverberation, sub-band synthesis, feature extraction and ASR, can now be jointly optimized using  the E2E ASR loss function. This model is refered to as E2E-DFAR model\footnote{\textcolor{black}{The implementation of the work can be found in \url{https://github.com/anurenjan/DFAR}}}.  The trainable  components are the DPLSTM model and the ASR model parameters, while the sub-band synthesis and feature extraction parameters are not learnable. 
% \begin{figure*}[t!]
%   \centering
%   \includegraphics[width=0.8\textwidth]{figs/carrier_plot_trial_3.pdf}

%       \caption{Comparison of mel-spectrograms before and after dereverberation (last two plots) for reverberant speech (far-room) recordings from the REVERB challenge dataset. The clean spectrogram is also shown at the top for reference.}

%   \label{fig:Spectrogram_compare}
% \end{figure*}

% \subsection{Visualization}
% Fig.~\ref{fig:Spectrogram_compare} shows the clean, reverberant and dereverberated mel spectrograms of an utterance from the REVERB Challenge   dataset (far-room). The reverberation effects are visible in the plot depicted in the second panel, where the temporal smearing blurs the details in the spectrogram. The dereverberated spectrogram, shown in the bottom panel, is able to retrieve some of the finer spectral details especially in low frequency regions. Experimental results shows that, these details are useful for in improving the ASR performance and for restoring the speech quality.

\section{Experimental setup}\label{sec:exp_set_up}
\subsection{Datasets}
\subsubsection{REVERB Challenge ASR}
The audio samples in REVERB challenge dataset \cite{rev3} are $8$ channel recordings with both real and simulated reverberant conditions. The real samples are utterances from MC-WSJ-AV corpus \cite{rev2}, spoken by human speakers in a noisy reverberant room. The simulated samples of the dataset are generated by convolving six different room impulse responses with the clean WSJCAM0 recordings followed by the addition of noise at the signal-to-noise ratio (SNR) of $20$ dB. The training data consists of $7861$ ( $\sim$ $17.5$ hours) utterances which are obtained by convolving WSJCAM0 train data with $24$ measured RIRs. The reverberation time of the measured impulse responses range from $0.2$ to $0.8$ sec. The training, development and evaluation data sets consist of $92$, $15$ and $38$  speakers respectively. The development data consists of $1663$  ($3.3$ hours) utterances and the evaluation data consists of $2548$  ($5.4$ hours) utterances. 

\subsubsection{VOiCES Dataset}
%\subsubsection{Data}
The VOiCES training set is a subset ($80$ hours) of the  LibriSpeech dataset. This set has utterances from $427$ speakers recorded in clean environments with close-talking microphones. The development and evaluation sets are far-field microphone recordings from diverse room dimensions, environments and noise conditions containing $19$ and $20$ hours of speech, respectively. The three sets namely training, development and evaluation, do not have any overlap in terms of the speakers. The robustness of the developed models is challenged by the mismatch that exists between the training and development/evaluation sets. We artificially added reverberation and noise on the $80$ hours training set, which served as the training set for all the E2E ASR experiments on the VOiCES dataset. The development set contains $20$ hours of distant recordings from the $200$  speakers. The evaluation data of $19$ hours consists of recordings $100$ speakers. The training set has $22741$ utterances, development set has $4318$ utterances and evaluation set has $4600$ utterances. 
 \input{Tables/Table_1_wer_rev} 

\subsection{E2E ASR baseline system}
For all the ASR experiments, we use the weighted prediction error based pre-processing~\cite{wpe} and unsupervised generalized eigenvalue (GEV) beamforming~\cite{rohit}. The baseline features are $36$-dimensional log-mel filter bank features with frequency range from $200$ Hz to $6500$ Hz. The ESPnet toolkit \cite{watanabe2018espnet} is used to perform all the end-to-end ASR experiments, with a Pytorch backend~\cite{pytorch}. The model architecture uses 12 conformer encoder layers with $2048$ units in the projection layer. The $6$-layer transformer architecture with $2048$ units in the  projection layer serves as the decoder. Both connectionist temporal cost (CTC) loss and attention based cross entropy (CE) loss are used in the training, with CTC-weight set at $0.3$ \cite{karita2019comparative}. A single layer of $1000$ LSTM cell recurrent neural network is used for  language modeling (RNN-LM). For training the model, we use stochastic gradient descent (SGD) optimizer with a batch size of $32$. For language model training, data is augmented from Wall Street Journal (WSJ)  corpus.

\subsection{Performance metrics}
\subsubsection{ASR performance metrics}
\begin{itemize}
    \item \textbf{WER/CER} (Word/Character Error Rate): The word/character error rate is given by the ratio of number of word/character insertions, deletions and substitutions in the system output to the total number of words/characters in the reference.
\end{itemize}

\subsubsection{Speech quality metrics}
\begin{itemize}
\item {\textbf{SRMR}}:
Speech to reverberation modulation ratio (SRMR) is a non intrusive measure. Here, a  representation is obtained
using an auditory-inspired filter bank analysis of critical band temporal envelopes of the signal. The modulation spectral
information is used to get an adaptive measure termed as speech to reverberation modulation energy ratio\cite{srmr_0, srmr}. A higher value indicates an  improved quality of the given speech signal.

%\item{\textbf{PESQ} (Perceptual Evaluation of Speech Quality)}: 
%Perceptual Evaluation of Speech Quality (PESQ) is an intrusive measure, which requires a clean reference signal. The technique was developed to model subjective tests commonly used in telecommunications \cite{pesq}. 
%The PESQ is the result of integration of  perceptual analysis measurement
%system (PAMS) \cite{pam2, pam1} and PSQM99 \cite{psqm1} which can be used over a wider variety of distortions and network conditions including packet loss, codecs, variable delay etc.
%The PESQ score is generated in the range of $1$-$4.5$, where the attempt is to model the $5$ point subjective scores. 

\item \textbf{MOS} (Mean Opinion Score): To evaluate the performance of
dereverberation algorithms, subjective quality 
and intelligibility measurement methods are needed. The most widely used subjective method is the ITU-T standard  \cite{recommendation2018itu}, where a panel of listeners are asked to rate the quality/intelligibility of the audio.  %Commonly, subjective quality tests have listeners rate
% the quality of the speech signal on a pre-specified scale.
    
\end{itemize}

\section{Experiments and results}~\label{sec:expt}
The baseline features are the beamformed log-mel filter-bank energy features (denoted as BF-FBANK).

\subsection{REVERB Challenge ASR}
% The REVERB challenge dataset \cite{reverb} for ASR consists of $8$ channel recordings with real and simulated reverberation conditions. The  simulated data is comprised of reverberant utterances (from the WSJCAM0 corpus \cite{rev1}) obtained by artificially convolving clean WSJCAM0 recordings with the measured room impulse responses (RIRs) and adding noise at an SNR of $20$ dB. The simulated data has six different reverberation conditions. The real data, which is comprised of utterances from the MC-WSJ-AV corpus \cite{rev2}, consists of utterances spoken by human speakers in a noisy reverberant room. The training set consists of $7861$ utterances from the clean WSJCAM0 training data convolved with $24$ measured RIRs.

\input{Tables/Table_4_wer_rvb_comp}
%\subsubsection{Discussion}
The word error rates (WER) for the dereverberation experiments are shown in Table \ref{table:1}. 
Note that, all the experiments use the same input features (log-mel filter bank features) along with the same E2E ASR architecture (conformer encoder and transformer decoder). 
The only difference between the various rows, reported in Table~\ref{table:1}, is the dereverberation pre-processing applied on the raw audio waveform. All the dereverberation experiments use the DPLSTM architecture described in Sec.~\ref{sec:proposed model}. 
\input{Tables/Table_0_wer_archi_trial}
\subsubsection{Various dereverberation configurations}
In Table~\ref{table:1}, the first row is the baseline result with the beamformed audio (unsupervised GEV beamforming \cite{rohit} and weighted prediction error (WPE) processing \cite{wpe}.

\textcolor{black}{The next set of rows compare several prior works. 
\begin{itemize}
    \item Fullsubnet - A full-band and sub-band fusion model for speech enhancement \cite{hao2021fullsubnet}.
    \item DCCRN - Deep complex convolution recurrent neural network  model for speech enhancement \cite{hu2020dccrn}. 
    \item Deep non-linear filter for multi-channel audio \cite{tesch2022insights}
    \item Reverberation time shortening \cite{zhou2023speech}
\end{itemize}
\textcolor{black}{The prior works are trained on the same data settings as used in the DFAR framework.}
All the prior works, except DCCRN (which is not designed for ASR), improve the baseline system in range of $8$-$11$\% in terms of relative WER. However, the proposed DFAR/E2E-DFAR approach is observed to provide the best WER, with relative improvement in WER of $19$/$34$\% on the evaluation data.}

\textcolor{black}{In Table~\ref{table:1}, we have also  performed two ASR experiments - i) using STFT inputs (log magnitude), and ii) using the sub-band signal directly without the envelope-carrier decomposition. Both these experiments, use the DPLSTM dereverberation model  proposed in this work. 
As seen in Table~\ref{table:1}, the dereverberation on the STFT magnitude component improves the ASR systems significantly over the baseline, while the dereverberation on the sub-band signal directly is not effective. However, the STFT approach is also seen to be inferior to the DFAR approach where the envelope-carrier dereverberation is performed.}

The fourth set of rows corresponds to the WER results with envelope/carrier based dereverberation alone. The relative improvements of $2-9$\% are seen here compared to the baseline BF-FBANK. Separately, with dereverberation based on the carrier signal alone,  a similar improvement is achieved. Further, the dereverberation of the temporal envelope and carrier components in a combined fashion using the DPLSTM model improves the ASR results over the separate dereverberation of envelope/carrier components. Here,  average relative improvements of $16$\% and $19$\% are seen in the development set and evaluation set respectively, over the BF-FBANK baseline system for the DFAR approach.

The final row in Table~\ref{table:1} reports the results using the joint learning of the dereverberation network and the E2E ASR model.  The E2E-DFAR is initialized using the dereverberation model and the E2E model trained separately.   The proposed E2E-DFAR model yields average relative improvements of $27$\% and $34$\% on the development set and evaluation set respectively over the baseline system. The joint training is also shown to improve over the set up of having separate networks for dereverberation and E2E ASR. 
While the DFAR model is trained only on simulated reverberation conditions, the WER improvement in real condition is  seen to be  more pronounced than those observed in the simulated data. This indicates that the model can generalize well to unseen reverberation conditions in the real-world.

\input{Tables/Table_6_hyper_parameter_env_crr}

\input{Tables/Table_5_wer_voices}

\input{Tables/Table_2_srmr_pesq}
\input{Tables/Table_3_reverb_MOS}
\subsubsection{Comparison with prior works}

The comparison of the results from prior works reported on the REVERB challenge dataset is given in Table~\ref{table:reverb_Prior}. The Table includes results from end-to-end ASR systems~\cite{subramanian2019investigation,zhang2020end,fujita2020attention} as well as the joint enhancement and ASR modeling work reported in \cite{heymann2019joint}. 
We also compare with our prior work reported in \cite{purushothaman2022dereverberation}. 
\textcolor{black}{
Specifically, many of the prior works compared in Table~\ref{table:reverb_Prior} are based on STFT based enhancement. The work reported in Subramanian et. al. \cite{subramanian2019investigation}, used a neural beamforming approach in the STFT domain, while the  efforts described in Heymann et. al. \cite{heymann2019joint}, used a long-short term memory network for mask estimation in power spectral domain (PSD). The dynamic convolution method proposed in Fujita et. al. \cite{fujita2020attention} used deconvolution of log-mel spectrogram features. Similar to the proposed work, all these efforts have also used the E2E ASR model training. As seen in Table \ref{table:reverb_Prior}, the proposed work improves over  these prior works considered here, further highlighting the benefits of the dereverberation in the sub-band time domain using long-term envelope-carrier based DPLSTM models. }

% To the best of our knowledge, the results from the proposed E2E-DFAR constitute the best published ASR  performance on the REVERB challenge evaluation dataset (relative improvements of $12$\% over the recent work by Fujita et. al.~\cite{fujita2020attention}). 

\subsubsection{Dereverberation model architecture}

The ASR experiments on the REVERB challenge dataset, pertaining to the choice of different model architectures used in the dereverberation model, are listed in Table~\ref{table:6}. \textcolor{black}{We have experimented with convolutional LSTM (CLSTM) \cite{kumar2022end} and time-domain LSTM (4-layer LSTM) architecture \cite{mimura2015speech} in addition to the DPLSTM approach}. As seen here, the Dual-path recurrence based DPLSTM gives the best word error rate in comparison with the other LSTM neural architectures considered. This may be attributed to the joint time-frequency recurrence performed to the other approaches which perform only time domain recurrence. 

\subsubsection{Dereverberation loss function}

The MSE loss function used in the DPLSTM model training consists of a combination of loss values from the envelope and the carrier components.  We experimented with the hyper parameter, $\lambda$, which controls the proportion of envelope based loss and carrier based loss in the total loss ($Total~loss = \lambda \times env.~ loss + (1-\lambda) \times carr.~ loss$). The ASR results for the various choices of the hyper parameter $\lambda$ are shown in Table~\ref{table:7}. Empirically, the value of $\lambda = 0.6$ gives the best WER on the REVERB challenge dataset. \textcolor{black}{Further, the choice of $\lambda = 1$ or $\lambda = 0$}, corresponding to envelope/carrier only dereverberation, are inferior to other choices of $\lambda$, indicating that the joint dereverberation of  the envelope and carrier components is beneficial.

\subsection{VOiCES ASR}
% The training set of the VOiCES corpus \cite{voices} consists of $80$-hour subset of the clean LibriSpeech corpus. The training set has close talking microphone recordings from $427$ speakers recorded in clean environments. The development and evaluation set consists of $19$ hours and $20$ hours of far-field microphone recordings from diverse room dimensions, environment and noise conditions. There are no common speakers between the training set and the development set or the  evaluation set. The significant  mismatch the training set and development/evaluation set allows the testing of the robustness of the trained models. We have used the same transformer based E2E ASR system that was developed for the REVERB challenge dataset. Further, the current experiments do not perform any data augmentation in the ASR model training. 

%\subsubsection{Discussion}
 The ASR setup used in the VOiCES dataset followed the ESPnet recipe with the conformer encoder and a transformer decoder. The rest of the model parameters and hyper-parameters are kept similar to the ones in the REVERB challenge dataset. The WER results on the VOiCES dataset are given in Table \ref{table:voice_table}. The dereverberation of the envelope alone provides an absolute improvement of $1.9$\% and $2.2$\% on the development and evaluation data respectively, compared to the FBANK baseline system. The dereverberation based on envelope-carrier modeling further improves the results. An absolute improvement of $3.3$\%/$5.4$\% on the development/evaluation data is achieved, compared to the FBANK baseline. Further, the joint training on envelope-carrier dereverberation network with the ASR model improves the WER results. We observe relative improvements of $10$\% and $12$\% on the development and evaluation data respectively . 
 
\subsection{Speech quality evaluation}
A comparison of the SRMR values for different dereverberation approaches is reported in Table \ref{table:2}. Here, we compare the baseline unsupervised GEV beamforming \cite{rohit} and weighted prediction error (WPE) \cite{wpe} with various strategies for beamforming. The deep complex convolutional recurrent network (DCCRN) based speech enhancement \cite{hu2020dccrn} is also implemented on the REVERB dataset, and these results are reported in Table  \ref{table:2}. While the envelope based dereverberation did not improve the SRMR values, the carrier based dereverberation is shown to improve the SRMR results. Further, the DFAR model also achieves similar improvements in SRMR for all the conditions over the baseline approach (GEV+WPE) and the DCCRN approach.

We conducted a subjective evaluation to further assess the performance of the dereverberation method. The subjects were asked to rate the quality of the audio on a scale of $1$ to $5$, $1$ being poor and $5$ being excellent. The subjects listened to the audio in a relatively quiet room with a high quality Sennheiser headset. We perform the A-B listening test, where the two versions of the same audio file were played, the first one with GEV + WPE dereverberation and the second one  with the proposed dereverberation approach. 
We chose $20$ audio samples, from four different conditions (real and simulated data and from near and far rooms) for this evaluation and recruited $20$ subjects. 

The subjective results are shown in Table \ref{table:3}. As seen, the proposed speech dereverberation scheme shows improvement in subjective MOS scores for all the conditions considered. 
The subjective results validate the signal quality improvements observed in the SRMR values (Table~\ref{table:2}).

\section{Conclusion}\label{sec:summary}
In this paper, we propose a speech dereverberation model using frequency domain linear prediction based sub-band envelope-carrier decomposition. The sub-band envelope and carrier components are processed through a dereverberation network. A novel neural architecture, based on dual path recurrence, is proposed for dereverberation. Using the joint learning of the neural speech dereverberation module and the E2E ASR model, we perform several speech recognition experiments on the  REVERB challenge dataset as well as on the VOiCES dataset. These results show that the proposed approach improves over the state of art E2E ASR systems based on mel filterbank features. 

The dereverberation approach proposed in this paper also  reconstructs the audio signal, which makes it useful for audio quality improvement applications as well as other speech processing systems in addition to the ASR system. We have further evaluated the reconstruction quality subjectively  and objectively   on the REVERB challenge dataset. The quality measurements show that the proposed speech dereverberation method improves speech quality over the baseline framework of weighted prediction error. The ablation studies on various architecture choices provides justification for the choice of the DPLSTM network architecture. 
Given that the proposed model allows the reconstruction of the audio signal, it can be used in conjunction with self-supervised neural approaches for representation learning of speech as well. This will form part of our future investigation.
\IEEEpubidadjcol

\bibliographystyle{IEEEtran}
\bibliography{taslp}

% \newpage

\begin{IEEEbiography}[{\includegraphics[width=1in,height=1.25in,clip,keepaspectratio]{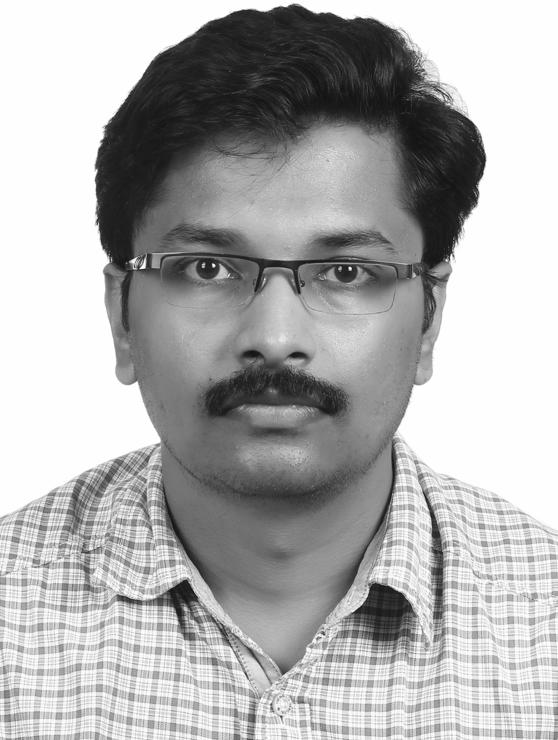}}]{Anurenjan Purushothaman}
is associated with the Learning and Extraction
of Acoustic Patterns (LEAP) lab, Department of Electrical Engineering, Indian Institute of Science, Bangalore, India, 56001, where he is pursuing his PhD under the guidance of Prof. Sriram Ganapathy. He is also associated with College of Engineering, Trivandrum and Government Engineering College, Idukki, where he is an Assistant Professor in the Department of Electronics \& Communication. He received the Bachelor's degree in Electronics \& Communication Engineering from Govt. Engineering College, Barton Hill, Trivandrum in 2006, the Master's degree in Signal Processing from College of Engineering, Trivandrum in 2008. In March 2008 he joined the Department of ECE,  College of Engineering,  Trivandrum as Assistant Professor.
\end{IEEEbiography}

\begin{IEEEbiography}[{\includegraphics[width=1in,height=1.25in,clip,keepaspectratio]{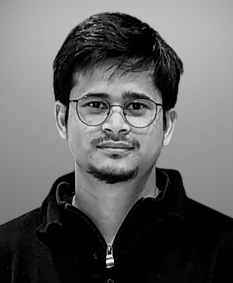}}]{Debottam Dutta} is currently a PhD student at the Signals \& Inference Research Group (SiNRG) at University of Illinois at Urbana-Champaign (UIUC). Prior to joining UIUC, he worked as a Senior Research Fellow for a year at the Learning and Extraction of Acoustic Patterns (LEAP) lab, Indian Institute of Science, Bangalore. He obtained his Master of Technology degree in Signal Processing from Indian Institute of Science, Bangalore in 2021 and Bachelor's degree in Electronics and Communication Engineering from National Institute of Technology, Silchar in 2018. 
\end{IEEEbiography}
\begin{IEEEbiography}[{\includegraphics[width=1in,height=1.25in,clip,keepaspectratio]{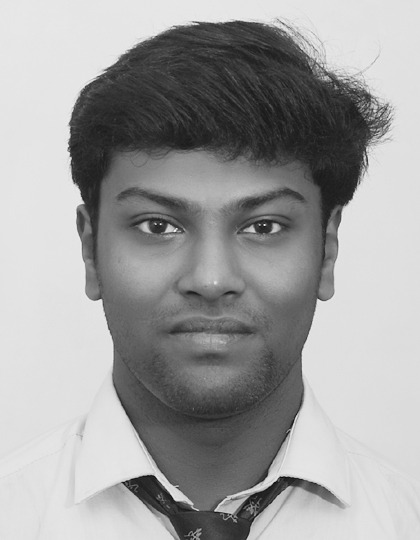}}]
{Rohit Kumar}
is currently associated with the Laboratory for Computational Audio Perception (LCAP) at Johns Hopkins University, Baltimore, Maryland. He is pursuing his PhD under the guidance of Prof. Mounya Elhilali. Rohit earned his Bachelor's degree in Electronics \& Communication Engineering from Delhi Technological University (formerly Delhi College of Engineering) in 2017 and his Master's degree in Signal Processing from the Indian Institute of Science, Bangalore, India, in 2020, under the supervision of Prof. Sriram Ganapathy. Prior to his doctoral studies, he worked as a research fellow in the Learning and Extraction of Acoustic Patterns (LEAP) lab at the Indian Institute of Science, Bangalore, India.
\end{IEEEbiography}
\begin{IEEEbiography}[{\includegraphics[width=1in,height=1.25in,clip,keepaspectratio]{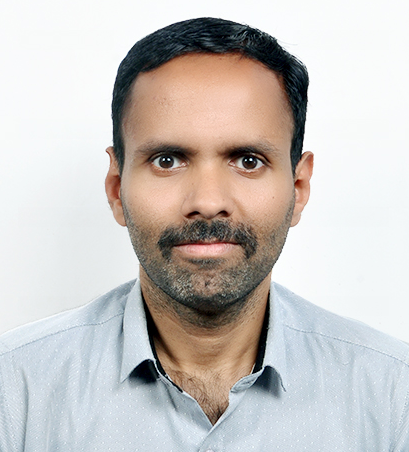}}]{Sriram Ganapathy}
is an Associate Professor at the Electrical Engineering, Indian Institute of Science, Bangalore, where he heads the activities of the learning and extraction of acoustic patterns (LEAP) lab. He is also associated with the Google Research India, Bangalore. Prior to joining the Indian Institute of Science, he was a research staff member at the IBM Watson Research Center, Yorktown Heights. He received his Doctor of Philosophy from the Center for Language and Speech Processing, Johns Hopkins University. He obtained his Bachelor of Technology from College of Engineering, Trivandrum, India and Master of Engineering from the Indian Institute of Science, Bangalore. He has also worked as a Research Assistant in Idiap Research Institute, Switzerland from 2006 to 2008.  At the LEAP lab,  his research interests include signal processing, machine learning methodologies for speech and speaker recognition and auditory neuro-science. He is a subject editor for the Speech Communications journal and a senior member of the IEEE. 
\end{IEEEbiography}
 

% \end{IEEEbiography}



\vfill

\end{document}

%% file: Tables/Table_1_wer_rev.tex
\begin{table*}[t!]
\caption{WER (\%) on the REVERB dataset for envelope/carrier, envelope-carrier dereverberation (DFAR) and the joint E2E-DFAR model. The relative improvements (\%) pertain to the comparison of the various configurations w.r.t. the BF-FBANK baseline system.}
\begin{center}
{
\centering
\resizebox{0.89\textwidth}{!}{%
\begin{tabular}{@{}l|lll|lll@{}}
\toprule
\multicolumn{1}{l|}{\multirow{2}{*}{\textbf{\begin{tabular}[c]{@{}l@{}}
Model\\ Config.
\end{tabular}}}} &   \multicolumn{3}{c|}{\textbf{Dev}}                                              & \multicolumn{3}{c}{\textbf{Eval}}                                             \\ \cmidrule(l){2-7} 
\multicolumn{1}{c|}{}                                                                                   &  \multicolumn{1}{l}{\textbf{Real}} & \multicolumn{1}{l}{\textbf{Sim.}} & \multicolumn{1}{l|}{\textbf{Avg. [Rel. Imp.]}} & \multicolumn{1}{l}{\textbf{Real}} & \multicolumn{1}{l}{\textbf{Sim}} & \multicolumn{1}{l}{\textbf{Avg. [Rel. Imp.]}} \\ \midrule
BF-FBANK  (baseline)  &  12.8                     & 8.7                      & 10.8 [- $ $-$ $ -] & 11.9                     & 7.9                      & 9.9 [- $ $-$ $ -]                   \\ \midrule 

\textcolor{black}{DCCRN\cite{hu2020dccrn} + BF-FBANK}  &   \textcolor{black}{ 17.4}                & \textcolor{black}{10.4}                   & \textcolor{black}{13.9 [-28.7]   }                  & \textcolor{black}{15.3}                     & \textcolor{black}{8.8}                    & \textcolor{black}{12.1 [-22.2]} \\
\textcolor{black}{Fullsubnet + BF-FBANK \cite{hao2021fullsubnet}}  &    \textcolor{black}{11.8}                   & \textcolor{black}{7.9}                     & \textcolor{black}{9.9 [+8.3]}                    & \textcolor{black}{10.5   }                  &\textcolor{black}{ 7.4 }                     &\textcolor{black}{ 9.0 [+9.1]} \\
\textcolor{black}{Deep non-linear filter\cite{tesch2022insights} + BF-FBANK}  &    \textcolor{black}{12.4}                    & \textcolor{black}{8.1}                    & \textcolor{black}{10.3 [+4.6] }                   & \textcolor{black}{10.5}                     & \textcolor{black}{7.2  }                    & \textcolor{black}{8.9 [+10.1]} \\
\textcolor{black}{Reverb. time shortening \cite{zhou2023speech}+ BF-FBANK}  &    \textcolor{black}{11.5}                   & \textcolor{black}{7.6}                     & \textcolor{black}{9.6 [+11.1]}                     & \textcolor{black}{10.1 }                    & \textcolor{black}{7.6 }                     & \textcolor{black}{8.9 [+10.1]} \\
\midrule
\textcolor{black}{ STFT Deverb. + BF-FBANK}  &    \textcolor{black}{12.0}                   & \textcolor{black}{7.8}                    & \textcolor{black}{9.9 [+8.3]}                     & \textcolor{black}{10.8}                    & \textcolor{black}{7.3}                      & \textcolor{black}{9.1 [+8.1] }\\
 \textcolor{black}{Sub-band sig. Dereverb. + BF-FBANK}  &   \textcolor{black}{13.3}                   & \textcolor{black}{9.8}                     & \textcolor{black}{11.6 [-7.4] }                    & \textcolor{black}{12.8}                     & \textcolor{black}{8.6}                      & \textcolor{black}{10.7 [-8.1]} \\ 
 \midrule 
FDLP Env. Derevb. + BF-FBANK    &    12.7                    & 8.5                     & 10.6 [+1.9]                     & 10.1                     & 7.8                      & 9.0 [+9.1] \\
FDLP Carr. Dereverb. + BF-FBANK & 11.2 & 8.3 & 9.8 [+9.3]                      & 10.8                     & 7.6                      & 9.2 [+7.1] \\ 
DFAR + BF-FBANK    & 10.6                    & 7.6                     & 9.1  [+15.7]                    & 9.1                     & 6.9                       & 8.0 [+19.2]                    \\ \midrule 
E2E-DFAR &                                                                                             \textbf{9.4  }                  & \textbf{6.4  }                   & \textbf{7.9 [+26.9]}                      & \textbf{7.3  }                   & \textbf{5.7 }                      & \textbf{6.5 [+34.3]} 
\\ \bottomrule

\end{tabular}}}

\label{table:1}
\end{center}
\vspace{-0.4cm}
\end{table*}

%% file: Tables/Table_4_wer_rvb_comp.tex
\begin{table}[t!]
\centering
\caption{Comparison of the results with other works reported on the REVERB challenge dataset. }
\resizebox{0.94\columnwidth}{!}{%
\begin{tabular}{l|c|c|c}
\toprule
\textbf{System} & \textbf{Eval-sim.} & \textbf{Eval-real} & \textbf{Avg.} \\   \midrule%\hline       
Subramanian et. al.~\cite{subramanian2019investigation} & 6.6 & 10.6 & 8.6 \\  
Heymann et. al.~\cite{heymann2019joint}      & -              & 10.8  & - \\    
Fujita et. al.~\cite{fujita2020attention} & \textbf{4.9 }& 9.8 & 7.4\\ 
Purushothaman et. al.~\cite{purushothaman2022dereverberation} & 7.1 & 12.1 & 9.6 \\ 
Zhang et. al.~\cite{zhang2020end}      & -              & 10.0  & - \\  \midrule  
This work & 5.7  & \textbf{7.3} &\textbf{ 6.5} \\
\bottomrule
\end{tabular}}
\vspace{-0.2cm}
\label{table:reverb_Prior}
\end{table}

%% file: Tables/Table_0_wer_archi_trial.tex
\begin{table}[t!]
\caption{WER (\%) in REVERB dataset for different architectures for the dereverberation model.}
\vspace{-0.4cm}
\begin{center}
{
\centering
\resizebox{0.9\columnwidth}{!}{%
\begin{tabular}{@{}l|ccc|ccc@{}}
\toprule
\multicolumn{1}{l|}{\multirow{2}{*}{\textbf{\begin{tabular}[c]{@{}c@{}}
Model\\ Config.
\end{tabular}}}} &   \multicolumn{3}{c|}{\textbf{Dev}}                                              & \multicolumn{3}{c}{\textbf{Eval}}                                             \\ \cmidrule(l){2-7} 
\multicolumn{1}{c|}{}                                                                                   &  \multicolumn{1}{l}{\textbf{Real}} & \multicolumn{1}{l}{\textbf{Sim}} & \multicolumn{1}{l|}{\textbf{Avg}} & \multicolumn{1}{l}{\textbf{Real}} & \multicolumn{1}{l}{\textbf{Sim}} & \multicolumn{1}{l}{\textbf{Avg}} \\ \midrule
Baseline  &  12.8                     & 8.7                      & 10.8 & 11.9                     & 7.9                      & 9.9                    \\

CLSTM   &    14.5                    & 9.7                     & 12.1                      & 12.4                     & 9.1                      & 10.8 \\ 
4-layer LSTM & 12.5 & 8.0 & 10.3                      & 10.1                     & 7.1                      & 8.9 \\ 
DPLSTM   & \textbf{10.6}                    & \textbf{7.6}                     & \textbf{9.1}                      & \textbf{9.1}                     & \textbf{6.9}                       & \textbf{8.0}                       
\\ \bottomrule

\end{tabular}}}

\label{table:6}
\end{center}
\vspace{-0.4cm}
\end{table}

%% file: Tables/Table_6_hyper_parameter_env_crr.tex
\begin{table}[t!]
\caption{WER (\%) in REVERB dataset for Hyper parameter $\lambda$, in $MSE~~loss = \lambda \times env.~ loss + (1-\lambda) \times carr.~ loss$.
}
\vspace{-0.4cm}
\begin{center}
{
\centering
\resizebox{0.9\columnwidth}{!}{%
\begin{tabular}{@{}c|ccc|ccc@{}}
\toprule
\multirow{2}{*}{\textbf{\begin{tabular}[c]{@{}c@{}}Parameter\\ $\lambda$\end{tabular}}} & \multicolumn{3}{c|}{\textbf{Dev}}            & \multicolumn{3}{c}{\textbf{Eval}}            \\ \cmidrule(l){2-7} 
                                                                                        & \textbf{Real} & \textbf{Simu} & \textbf{Avg} & \textbf{Real} & \textbf{Simu} & \textbf{Avg} \\ \midrule
0                                                                                       & 12            & 8.2           & 10.1         & 10.4          & 7.5           & 9.0            \\
0.2                                                                                     & 11.9          & 8.6           & 10.3         & 10.7          & 7.7           & 9.2          \\
0.4                                                                                     & 11.6          & 8.2           & 9.9          & 10.1          & 7.2           & 8.7          \\
0.5                                                                                     & 11.3          & \textbf{7.2}           & 9.3          & 9.7           & \textbf{6.5}           & 8.1          \\
0.6                                                                                     & \textbf{10.6}          & 7.6           & \textbf{9.1}          & \textbf{9.1}           & 6.9           & \textbf{8.0}            \\
0.8                                                                                     & 13.1          & 8.7           & 10.9         & 10.9          & 7.9           & 9.4          \\
1                                                                                       & 13.5          & 8.0             & 10.8         & 10.4          & 6.9           & 8.7          \\ \bottomrule
\end{tabular}}}

\label{table:7}
\end{center}
\vspace{-0.1in}
\end{table}

%% file: Tables/Table_5_wer_voices.tex
\begin{table}[t!]
\centering
\caption{Performance (WER \%) on the VOiCES dataset.}
\vspace{-0.2cm}
\resizebox{0.8\columnwidth}{!}{%
\begin{tabular}{l|c|c}
\toprule
\multirow{1}{*}{\textbf{Model Config.}} & \multicolumn{1}{c|}{\textbf{Dev}}            & \multicolumn{1}{c}{\textbf{Eval}}            
                                      \\ \midrule
FBANK   (baseline)                          &40.3              &50.8          \\
$~~~$ + Env. derevb.                             &38.4            & 48.6        \\
$~~~$ + Env.-carr. derevb. (DFAR)                       &37.1            & 45.4        \\ 
~~~ + E2E-DFAR                         & \textbf{36.4}           & \textbf{44.7}        \\
\bottomrule
\end{tabular}}
%\vspace{-0.5cm}
\label{table:voice_table}
\vspace{-0.1in}

\end{table}

%% file: Tables/Table_2_srmr_pesq.tex
\begin{table*}[t!]
\caption{SRMR values on the REVERB dataset for various signal enhancement strategies.}
\vspace{-0.4cm}
\begin{center}
{
\centering
\resizebox{0.95\textwidth}{!}{%
\begin{tabular}{l|cccccc}
\midrule
Signal                                          & \multicolumn{5}{c}{{SRMR}}                                                                                                                                                                                                                                                                                                                                                                                  \\ \midrule
 & Dev. (Real) 
 %\multicolumn{1}{c|}{\begin{tabular}[c]{@{}c@{}}Dev.\\ Real\end{tabular}}
 & Dev. (Sim.) & Eval. (Real) & Eval. (Sim.) & REVB. (Train) & \begin{tabular}[c]{@{}c@{}}\\ \end{tabular} \\ \midrule

\begin{tabular}[c]{@{}l@{}}Unsupervised GEV beamforming \cite{rohit} \end{tabular} & \multicolumn{1}{c|}{5.18}                                                & \multicolumn{1}{c|}{4.1}                                                 & \multicolumn{1}{c|}{4.58}                                                 & \multicolumn{1}{c|}{4.67}                                                 & 4.23             \\
\begin{tabular}[c]{@{}l@{}}$~~~$ + WPE \cite{wpe} \end{tabular} & \multicolumn{1}{c|}{5.35}                                                & \multicolumn{1}{c|}{4.2}                                                 & \multicolumn{1}{c|}{4.61}                                                 & \multicolumn{1}{c|}{4.75}                                                 & 4.48             \\ 
\begin{tabular}[c]{@{}l@{}} $~~~$ $~~~$ + DCCRN\cite{hu2020dccrn}\end{tabular} & \multicolumn{1}{c|}{5.43}                                                & \multicolumn{1}{c|}{4.37}                                                 & \multicolumn{1}{c|}{4.63}                                                 & \multicolumn{1}{c|}{4.94}                                                 & 4.67                                                              \\
\begin{tabular}[c]{@{}l@{}} $~~~$ $~~~$ \textcolor{black}{+} \textcolor{black}{Fullsubnet\cite{hao2021fullsubnet}}\end{tabular} & \multicolumn{1}{c|}{\textcolor{black}{5.36}}                                                & \multicolumn{1}{c|}{\textcolor{black}{4.32}}                                                 & \multicolumn{1}{c|}{\textcolor{black}{4.64}}                                                 & \multicolumn{1}{c|}{\textcolor{black}{4.97}}                                                 & \textcolor{black}{4.63}                                                              \\
\begin{tabular}[c]{@{}l@{}} $~~~$ $~~~$ \textcolor{black}{+ Deep Non-Linear Filters\cite{tesch2022insights}}\end{tabular} & \multicolumn{1}{c|}{\textcolor{black}{5.51}}                                                & \multicolumn{1}{c|}{\textcolor{black}{4.22}}                                                 & \multicolumn{1}{c|}{\textcolor{black}{4.64}}                                                 & \multicolumn{1}{c|}{\textcolor{black}{5.02}}                                                 & \textcolor{black}{4.61}                                                              \\
\begin{tabular}[c]{@{}l@{}} $~~~$ $~~~$ \textcolor{black}{+ Reverberation Time Shortening Target\cite{zhou2023speech}}\end{tabular} & \multicolumn{1}{c|}{\textcolor{black}{5.49}}                                                & \multicolumn{1}{c|}{\textbf{\textcolor{black}{4.57}}}                                                 & \multicolumn{1}{c|}{\textcolor{black}{4.62}}                                                 & \multicolumn{1}{c|}{\textcolor{black}{5.2}}                                                 & \textcolor{black}{4.58}                                                              \\
\begin{tabular}[c]{@{}l@{}} $~~~$ $~~~$ \textcolor{black}{+ STFT Mag. + DPLSTM}\end{tabular} & \multicolumn{1}{c|}{\textcolor{black}{5.44}}                                                & \multicolumn{1}{c|}{\textcolor{black}{4.33}}                                                 & \multicolumn{1}{c|}{\textcolor{black}{4.64}}                                                 & \multicolumn{1}{c|}{\textcolor{black}{4.94}}                                                 & \textcolor{black}{4.6}                                                            \\
\begin{tabular}[c]{@{}l@{}} $~~~$ $~~~$ \textcolor{black}{+ Sub-band signal + DPLSTM}\end{tabular} & \multicolumn{1}{c|}{\textcolor{black}{5.45}}                                                & \multicolumn{1}{c|}{\textcolor{black}{4.28}}                                                 & \multicolumn{1}{c|}{\textcolor{black}{4.61}}                                                 & \multicolumn{1}{c|}{\textcolor{black}{4.87}}                                                 & \textcolor{black}{4.63}                                                              \\
\begin{tabular}[c]{@{}l@{}} $~~~$ $~~~$ + env. derevb. (this work) \end{tabular} & \multicolumn{1}{c|}{4.62}                                                & \multicolumn{1}{c|}{3.83}                                                 & \multicolumn{1}{c|}{4.12}                                                 & \multicolumn{1}{c|}{4.25}                                                 & 4.11                                                                                                        \\ 
\begin{tabular}[c]{@{}l@{}}$~~~$ $~~~$ + crr. derevb. (this work) \end{tabular} & \multicolumn{1}{c|}{5.52}                                                & \multicolumn{1}{c|}{4.46}                                                 & \multicolumn{1}{c|}{4.69}                                                 & \multicolumn{1}{c|}{5.27}                                                 & 4.77                                                                                                        \\ 
\begin{tabular}[c]{@{}l@{}}$~~~$ $~~~$ + env. \& crr. derevb. [DFAR] (this work)\end{tabular} & \multicolumn{1}{c|}{\textbf{5.52}}                                                & \multicolumn{1}{c|}{4.47}                                                 & \multicolumn{1}{c|}{\textbf{4.69}}                                                 & \multicolumn{1}{c|}{\textbf{5.27}}                                                 & \textbf{4.77}                                                                                                                                                            \\ \midrule
\end{tabular}}}
\label{table:2}
\end{center}
\vspace{-0.4cm}
\end{table*}

%% file: Tables/Table_3_reverb_MOS.tex
\begin{table*}[t!]
\caption{MOS values in REVERB dataset for envelope and carrier based enhancements. }
%\vspace{-0.4cm}
\begin{center}
{
\centering
\resizebox{0.85\textwidth}{!}{%

\begin{tabular}{@{}l|c|c|c|c@{}}
\toprule
                     & \textbf{ET Real - near} & \textbf{ET Real - far} & \textbf{ET Simu - near} & \textbf{ET Simu - far} \\ \midrule
Baseline - GEV \cite{rohit} + WPE \cite{wpe} & 3.78                     & 3.65                    & 3.74                    & 4.12                   \\
$~~~$+ env.-carr. derevb. [DFAR] (this work)   & \textbf{3.98}                     & \textbf{3.67}                   & \textbf{4.01}                     & \textbf{4.40 }                  \\ \bottomrule
\end{tabular}
 }}
\label{table:3}
\end{center}
\vspace{-0.1in}

%\vspace{-0.4cm}
\end{table*}